\documentclass[preprint,11pt]{aastex}

\usepackage{graphics} 
\usepackage{graphicx}

\begin{document}

\slugcomment{To appear in {\em The Astrophysical Journal}}

\title{Structure and Magnetic Fields in the Precessing \\ Jet System SS\,433  I. Multi-Frequency Imaging from 1998}

\author{David H. Roberts, John F. C. Wardle, \\ Scott L. Lipnick, Philip L. Selesnick, and Simon Slutsky}
\affil{Department of Physics MS-057, Brandeis University, Waltham, MA 02454-0911 USA}
\email{roberts@brandeis.edu, wardle@brandeis.edu}

\begin{abstract}

The Very Large Array has been used at five frequencies to study the structure and linear polarization of SS\,433 on scales as small as $\sim$0.1$ \arcsec \simeq 500~\mbox{AU}$. Each jet consists of a sharp, curving ridge-line at the leading edge, plus significant trailing off-jet emission, showing that they are enveloped by diffuse relativistic plasma.  No kinematic model with constant jet speed fits our images on all scales, but they are consistent with variations in jet speed of around $\pm10\%$ around the optical value. Our images show continuous jets with bright components occurring simultaneously in the two jets roughly every 35~days. When corrected for projection effects and Doppler boosting, the intensities of the two jets are intrinsically very similar. Fractional linear polarization up to 20\% is present along the ridge-lines, while the core is essentially unpolarized. The rotation measures are consistent with a foreground screen with $\mbox{RM} \simeq +110$ rad m$^{-2}$ plus a larger, asymmetrical contribution close to the core. The apparent magnetic fields in the jets are roughly aligned with the ridge-lines in most but not all of each jet. The jet is more highly polarized between the components than in the components themselves, suggesting that the mechanism that creates them compresses a tangled part of the magnetic field into a partially-ordered transverse layer. The off-jet emission is remarkably highly polarized, with $m \simeq 50\%$ in places, suggesting large-scale order of the magnetic field surrounding the jets.  This polarized signal may confuse the determination of magnetic field orientations in the jets themselves. However, the images are consistent with a jet magnetic field that is everywhere parallel to the helices.
\end{abstract}


\section{INTRODUCTION}

The galactic binary source SS\,433, consisting of a stellar-mass black hole in close orbit about an early-type star, is the prototypical ``micro-quasar,'' a miniature version of an AGN \citep{MR99}. Two relativistic jets emerge from opposite sides of the compact object at speeds of $0.26~c$. Modeling of the optical spectrum shows that the jet system precesses with a period of 162 days about a cone of half-angle $20^{\circ}$ \citep{AM79, Fabian, Milgrom, M84}. Imaging by the Very Large Array (VLA) confirmed this picture, as the radio images showed helical jets on both sides of the source \citep{Spencer79,HJ81a,HJ81b}. Higher resolution images made by VLBI reveal the structure down to a scale of a few AU \citep{Ver87,Ver93}.

Arcsecond-scale imaging with the VLA at 5~GHz showed that the radio jets are linearly polarized at levels up to tens of percent, and that the polarized structure varies with position and precession phase \citep{HJ81a}. Hjellming and Johnston inferred that the underlying magnetic fields were roughly aligned
with ballistic (i.e., radial) motions in the jets, but acknowledged that in the absence of any detailed information on Faraday rotation in SS\,433, this conclusion was tentative. \cite{Gilmore81} also reported strong and variable polarization in the jets at two epochs. \cite{Stirl04} observed SS\,433 at multiple frequencies, and concluded that the jet magnetic fields lay along the spiral of the jet. At VLBI resolutions SS\,433 shows no linear polarization \citep{VLBP}. There is some circular polarization when observed at 1--9~GHz by the Australia Telescope Compact Array \citep{Fender2000}. 

In this paper we use the VLA to study the structure, linear polarization, and Faraday rotation of SS\,433 on scales as small as $\sim 0.1 \arcsec \simeq 500~\mbox{AU}$. In \S\ref{s:obs} we describe the observations and data reduction. In \S\ref{s:jets} we determine the properties of the jets, find the underlying magnetic field orientations, and discuss the physical implications of these results. Our conclusions are summarized in \S\ref{s:conclude}.

\section{OBSERVATIONS AND DATA REDUCTION}
\label{s:obs}

Interferometer data for three $\sim$4 hour periods over 1998 March 5--7 (MJD 10877--10879) were obtained from the VLA data archive. The array was in the A configuration, with 26 working antennas; the frequencies used were 1.415, 1.665, 4.800, and 14.94~GHz on March~5, and 1.415, 1.465, 4.800, and 14.94~GHz over March~6--7. The data were edited and phase calibrated in AIPS \citep{aips} using the nearby unresolved source J1950+081, and amplitude calibrated against 3C\,286. Instrumental polarization contamination was determined using the phase calibrator, and removed from the data in the standard manner, and electric vector position angles calibrated against 3C\,286. The data were then imaged in total intensity and linear polarization using DIFMAP \citep{difmap}, utilizing several cycles of phase and amplitude self-calibration. Natural weighting was used at 14.94~GHz, and uniform weighting at the lower frequencies. Images in Stokes $Q$ and $U$ were combined to make images of polarized intensity $p = \sqrt{Q^2 + U^2}$, electric vector polarization angle $\chi = (1/2) \arctan{(U/Q)}$, and fractional linear polarization $m = p/I$.

\begin{figure}[t]
\plottwo{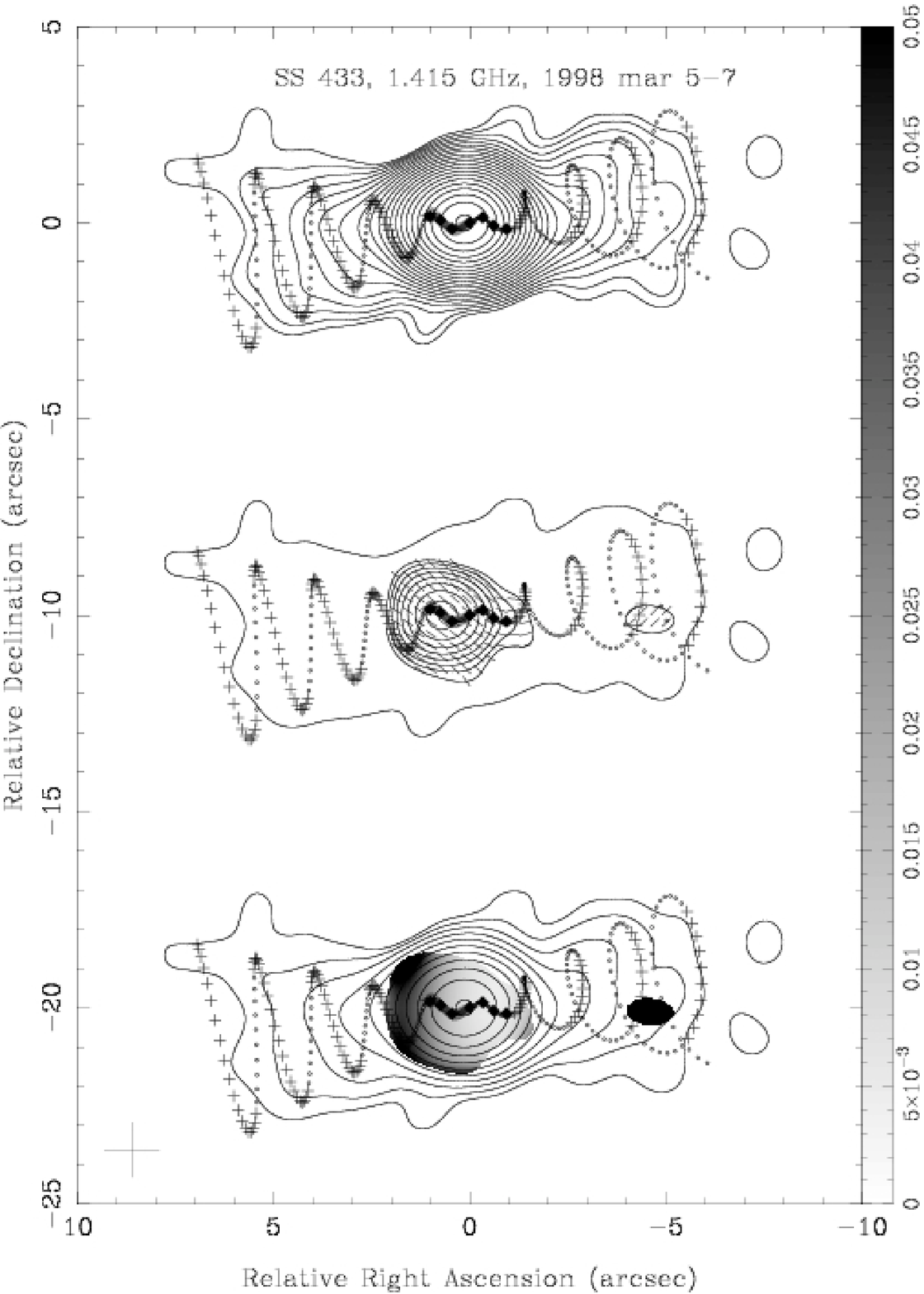}{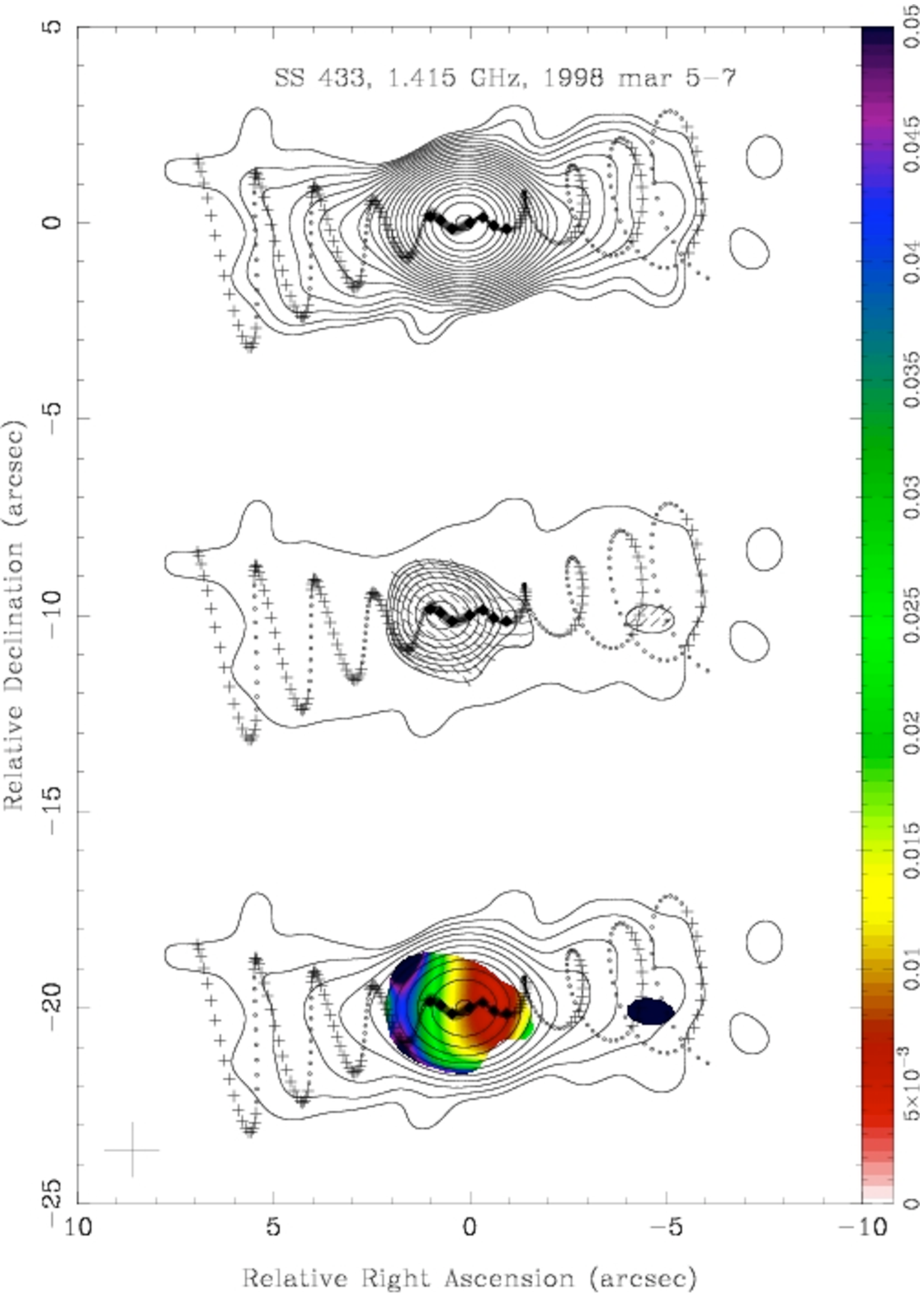}
\caption[]{VLA images of SS\,433 at 1.415~GHz, made from data taken over 1998 March 5-7. (Top) Contours of total intensity, starting at 0.75~mJy~beam$^{-1}$, with steps of factors of $\sqrt{2}$, and peak of 811~mJy~beam$^{-1}$. (Middle) Linear polarization, showing contours of polarized intensity, starting at 0.75~mJy~beam$^{-1}$, with steps of factors of $\sqrt{2}$, and peak of 10~mJy~beam$^{-1}$, with ticks indicating electric vector position angles, plus a single total intensity contour. (Bottom) Fractional polarization on the scale to the right, overlying every-other total intensity contour. The dots indicate the locations of seven bright total intensity components, as determined at 14.94~GHz (Figure~\ref{fig:U}). The kinematic model described in the text is shown as material ejected at five day intervals, with ``+''  and ``o'' indicating oncoming  and receding parts, respectively. The restoring beam with FWHM of  $1.38\arcsec \times 1.38\arcsec$ is shown as a cross to the lower-left. (Both monochrome and color figures are included for monochrome printers.)
\label{fig:L1}}
\end{figure}

\begin{figure}[t]
\plottwo{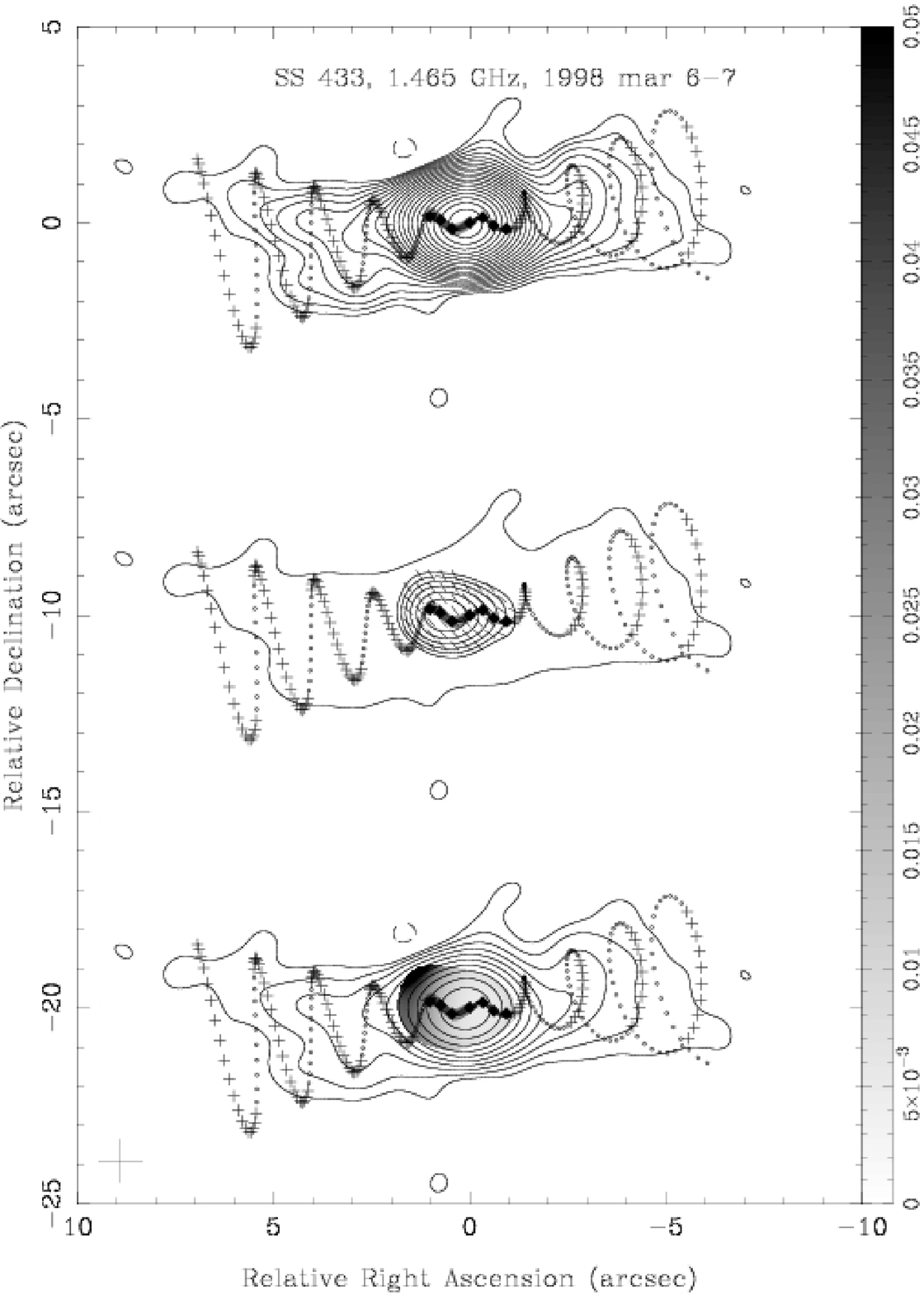}{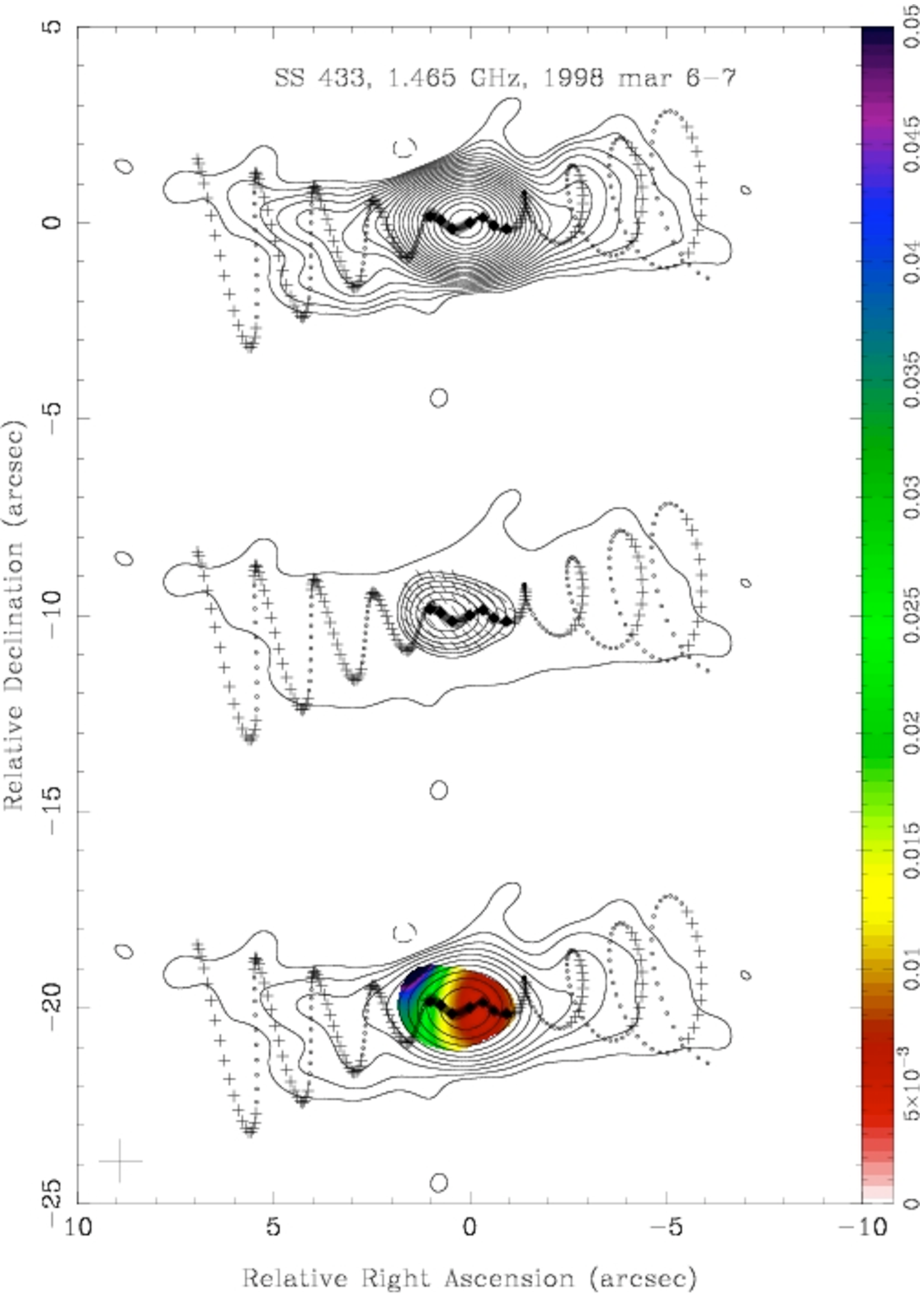}
\caption[]{VLA images of SS\,433 at 1.465~GHz, made from data taken over 1998 March 6-7. Same parts as Figure~\ref{fig:L1}. The contours of total intensity start at 0.75~mJy~beam$^{-1}$, have steps of factors of $\sqrt{2}$, and a peak of 732~mJy~beam$^{-1}$. The  contours of polarized intensity start at 0.75~mJy~beam$^{-1}$, have steps of factors of $\sqrt{2}$, and a peak of 5.9~mJy~beam$^{-1}$. The restoring beam with FWHM of  $1.10\arcsec \times 1.10\arcsec$ is shown as a cross to the lower-left. (Both monochrome and color figures are included for monochrome printers.)
\label{fig:L2}}
\end{figure}

\begin{figure}[t]
\plottwo{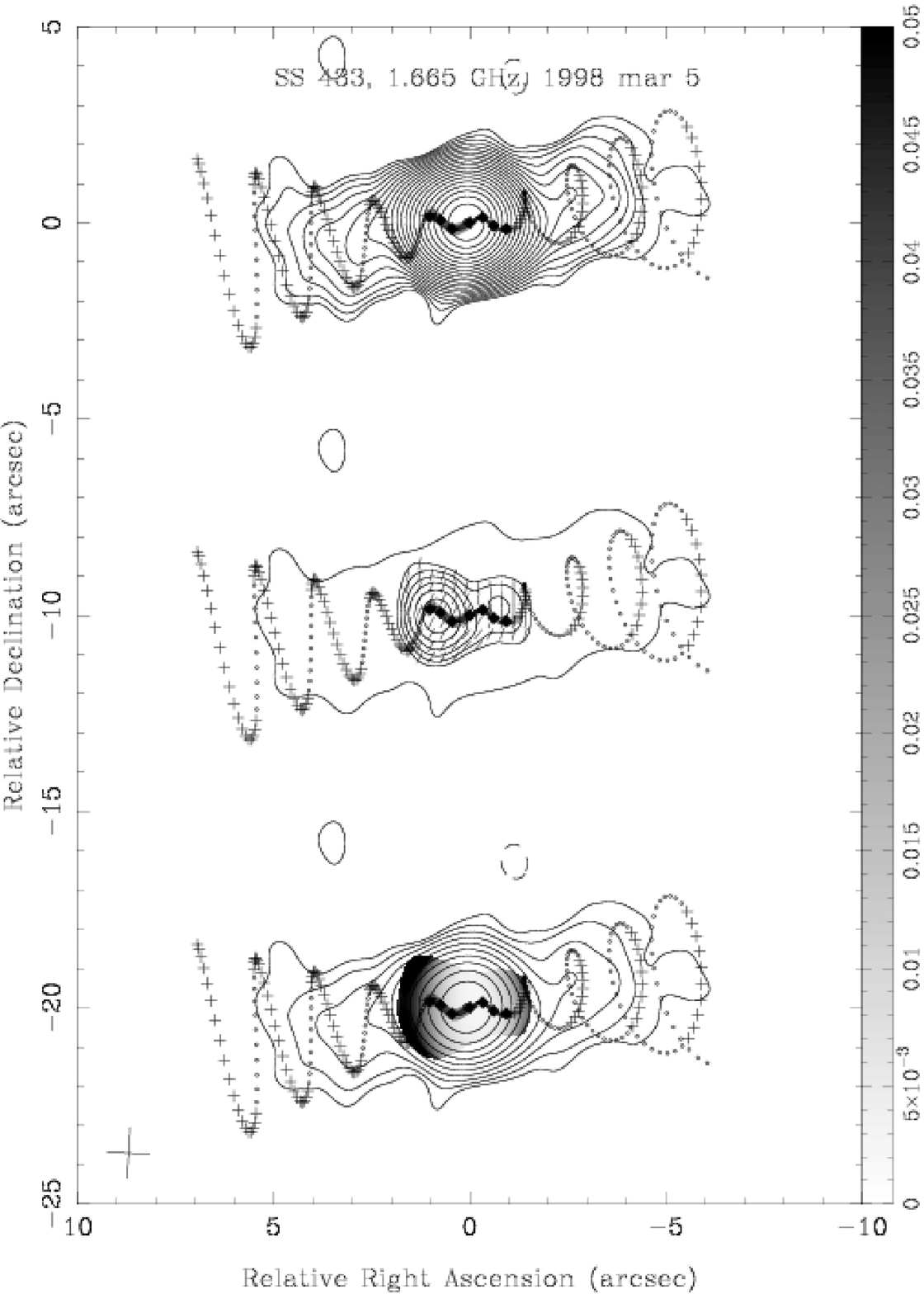}{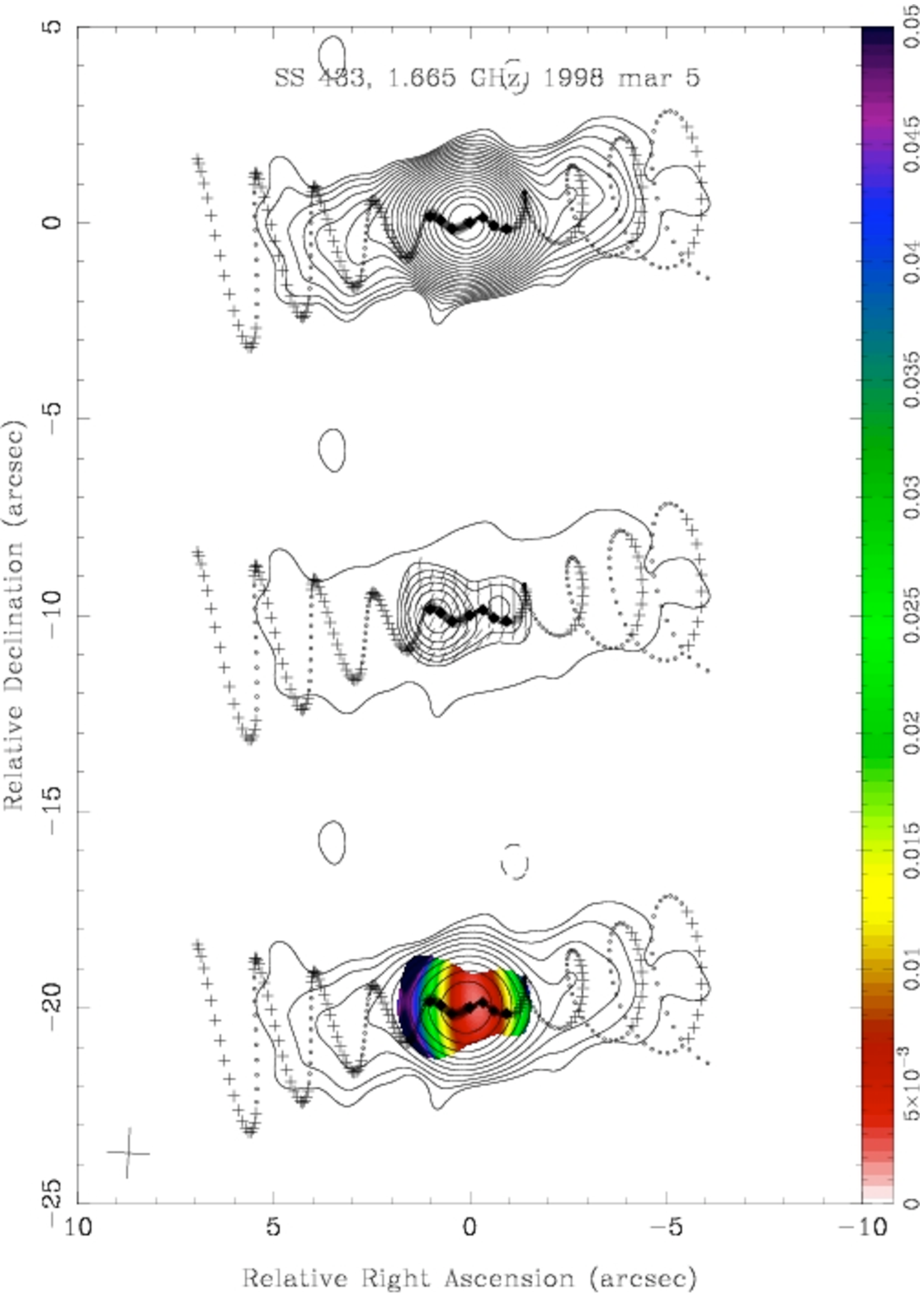}
\caption[]{VLA images of SS\,433 at 1.665~GHz, made from data taken 1998 March 5. Same parts as Figure~\ref{fig:L1}. The contours of total intensity start at 0.75~mJy~beam$^{-1}$, have steps of factors of $\sqrt{2}$, and a peak of 775~mJy~beam$^{-1}$. The  contours of polarized intensity start at 0.75~mJy~beam$^{-1}$, have steps of factors of $\sqrt{2}$, and a peak of 5.5~mJy~beam$^{-1}$. The restoring beam with FWHM of  $1.09\arcsec \times 1.09\arcsec$ is shown as a cross to the lower-left. (Both monochrome and color figures are included for monochrome printers.)
\label{fig:L3}}
\end{figure}

\begin{figure}[t]
\plottwo{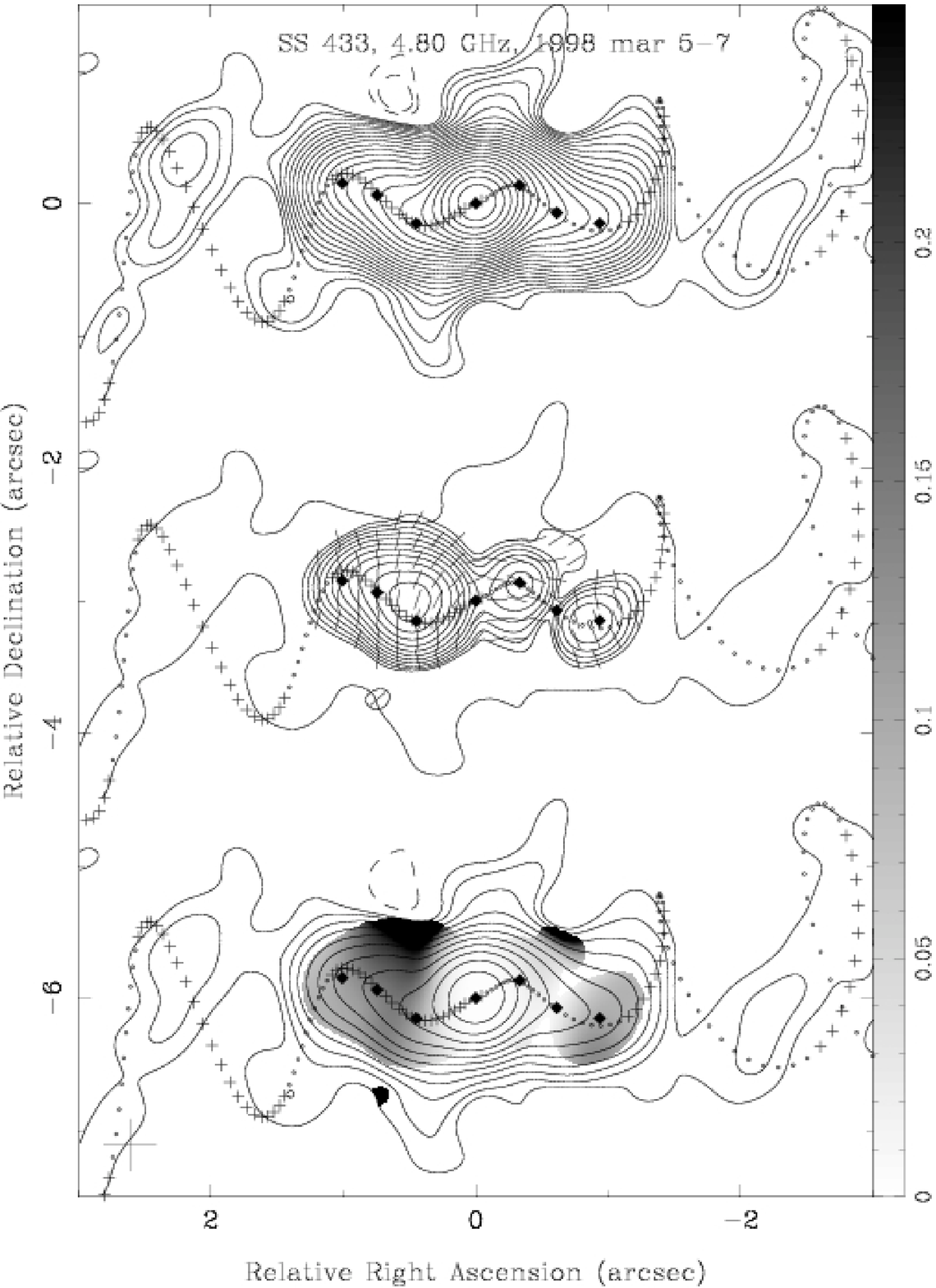}{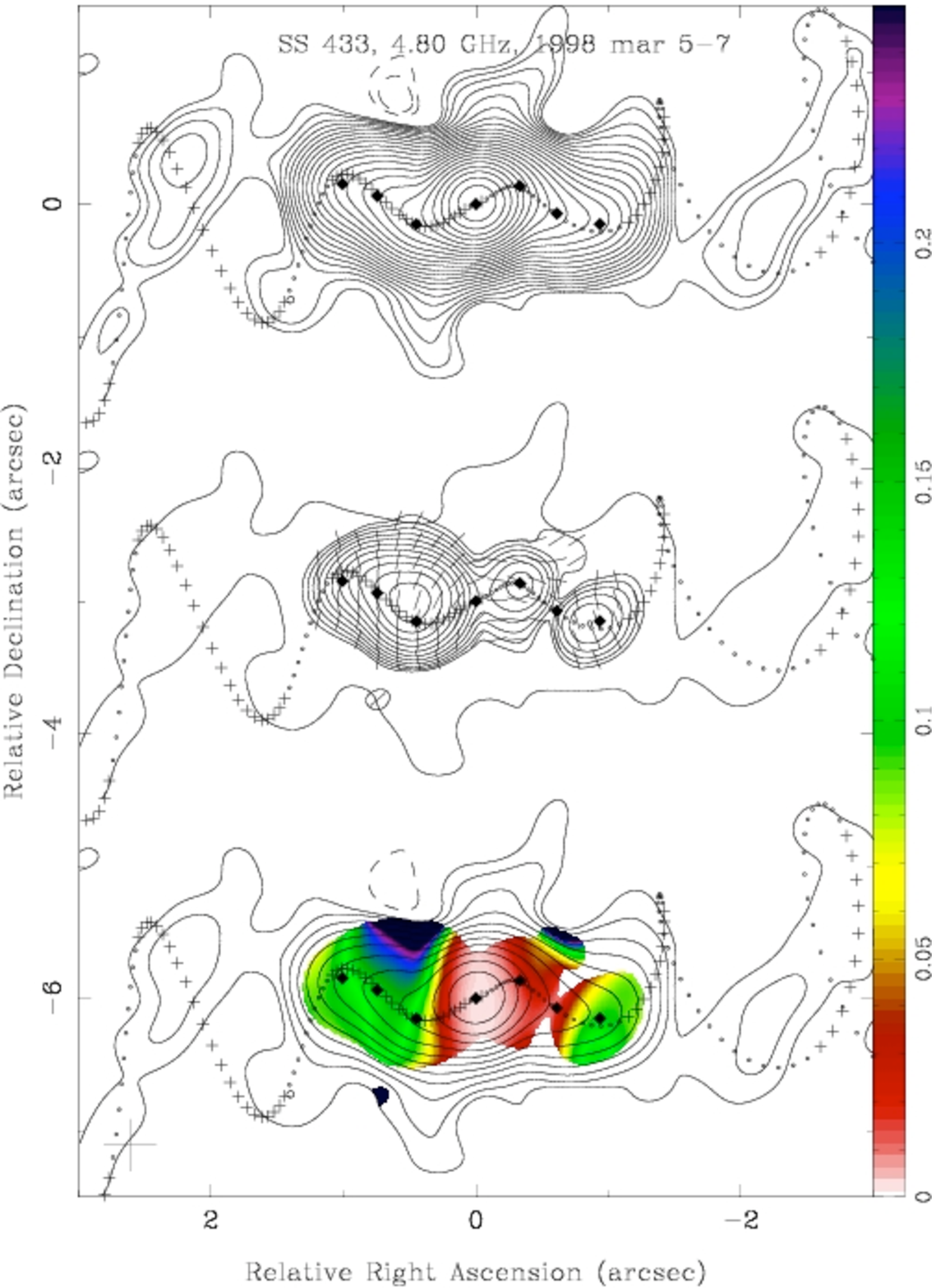}
\caption[]{VLA images of SS\,433 at 4.80~GHz, made from data taken 1998 March 5-7. Same parts as Figure~\ref{fig:L1}. The contours of total intensity start at 0.20~mJy~beam$^{-1}$, have steps of factors of $\sqrt{2}$, and a peak of 398~mJy~beam$^{-1}$. The  contours of polarized intensity start at 0.20~mJy~beam$^{-1}$, have steps of factors of $\sqrt{2}$, and a peak of 8.1~mJy~beam$^{-1}$. The restoring beam with FWHM of  $0.39\arcsec \times 0.39\arcsec$ is shown as a cross to the lower-left. (Both monochrome and color figures are included for monochrome printers.)
\label{fig:C}}
\end{figure}

Images were made at four frequencies for each of the three days, and they were consistent with each other, and with the small proper motions expected in the jets over this time interval ($\mu \simeq 9~\mbox{mas} \, \mbox{d}^{-1}$). The core varied by a few percent at every frequency at which we had more than one day's data. To optimize the signal-to-noise ratios in the jets, we aligned the images on the core, and simply averaged the multiple images at each of 1.415, 1.465, 4.800, and 14.94~GHz. Smearing of the jet features over the span of the observations amounts to only 15\% of a beamwidth at 15~GHz, and does not result in false structures, either in total intensity or in polarization. Along with the single image at 1.665~GHz, the results are presented in Figures~\ref{fig:L1}--\ref{fig:U}. We also created a 14.94~GHz image convolved to the resolution of the 4.80~GHz images (not shown). In each case the top panel shows contours of total intensity, the middle panel contours of linearly polarized intensity with electric vector position angle ticks, and the lower panel fractional polarization in grey scale (color in the electronic edition) over every-other total intensity contour. The images at 4.8 and 15~GHz are consistent with published images at different epochs and precession phases made by others (e.g., \citet{Stirl04, BB04}). In particular, the helical structure out to 3\arcsec\  from the core in Figure~\ref{fig:C} is real, and is not an artifact due to the sidelobe structure of the observations. This is supported by the agreement of the images and the kinematic model (see below).

On each image we have plotted model jets, with positions calculated from the kinematic model of Hjellming and Johnston (1981b), in which material is emitted as identical oppositely-directed narrow jets. In the convention of \cite{Ver93} (phase is zero when the east jet has maximum redshift), the precession phase at this epoch is $\phi=0.89$. Shown are predicted positions of material emitted at five-day intervals; material that is oncoming is marked with ``+,'' while retreating material indicated with ``o.''  We adopted the optical ephemeris of \cite{Eik}, combined with a position angle of the jet precession axis of $\mbox{PA} =  98.2^{\circ}$ determined from a sequence of VLBA images made in the summer of 2003 \citep{Modu}; see Table~\ref{tab:ephem}. Several distances have been proposed for SS\,433. \cite{Modu} found a distance of $d =  5.0~\mbox{kpc}$, \citet{Ver93} concluded from VLBI imaging that $d = 4.85 \pm 0.2 \, \mbox{kpc}$, \cite{Dubner} derived a distance of 3~kpc from \ion{H}{1} absorption, while \cite{Stirl02} found a distance of 4.8~kpc, with jet deceleration at the rate of $-0.02c$ per precession period.  \citet{BB04} concluded from a deep VLA image at 5~GHz that $d = 5.5 \, \mbox{kpc}$. Recently, \cite{LBG07} have re-examined the H~I distance, and conclude that it is in agreement with the kinematic distance of \cite{BB04}. In Figure~\ref{fig:CU3models} we plot our 4.80 and 14.94~GHz images against three kinematic models, (a) one with $d=5.0~\mbox{kpc}$ and $v=0.2647c$,  (b) one with $d=5.5~\mbox{kpc}$ and $v=0.2647c$, and (c) one with $d=5.5~\mbox{kpc}$ and $v=0.29c$. The fit to the inner parts of both jets at 14.94~GHz seems to favor either (a) or (c), but model (b) fits the larger-scale 4.80~GHz image. In accord with the conclusions of \citet{BB04} and \citet{LBG07}, we adopt $d=5.5$~kpc, and interpret the discrepancy in fitting the 14.94~GHz images as a sign that synchronized variations in velocities of about $\pm 10\%$ take place in the two jets, as suggested by \cite{Stirl04}, \cite{BB04,BB05}, and \cite{BBS07}. Since we are interested primarily in the smaller scale structure, in all figures we use $d=5.5~\mbox{kpc}$ and $v = 0.29c$. We also ignore the $6.5 \, \mbox{d}$ nodding motion \citep{Katz82} because it is small on the scales we are considering.

\section{PROPERTIES OF THE SS\,433 JETS}
\label{s:jets}

\subsection{Structure}
\label{s:structure}

\begin{figure}[t]
\plottwo{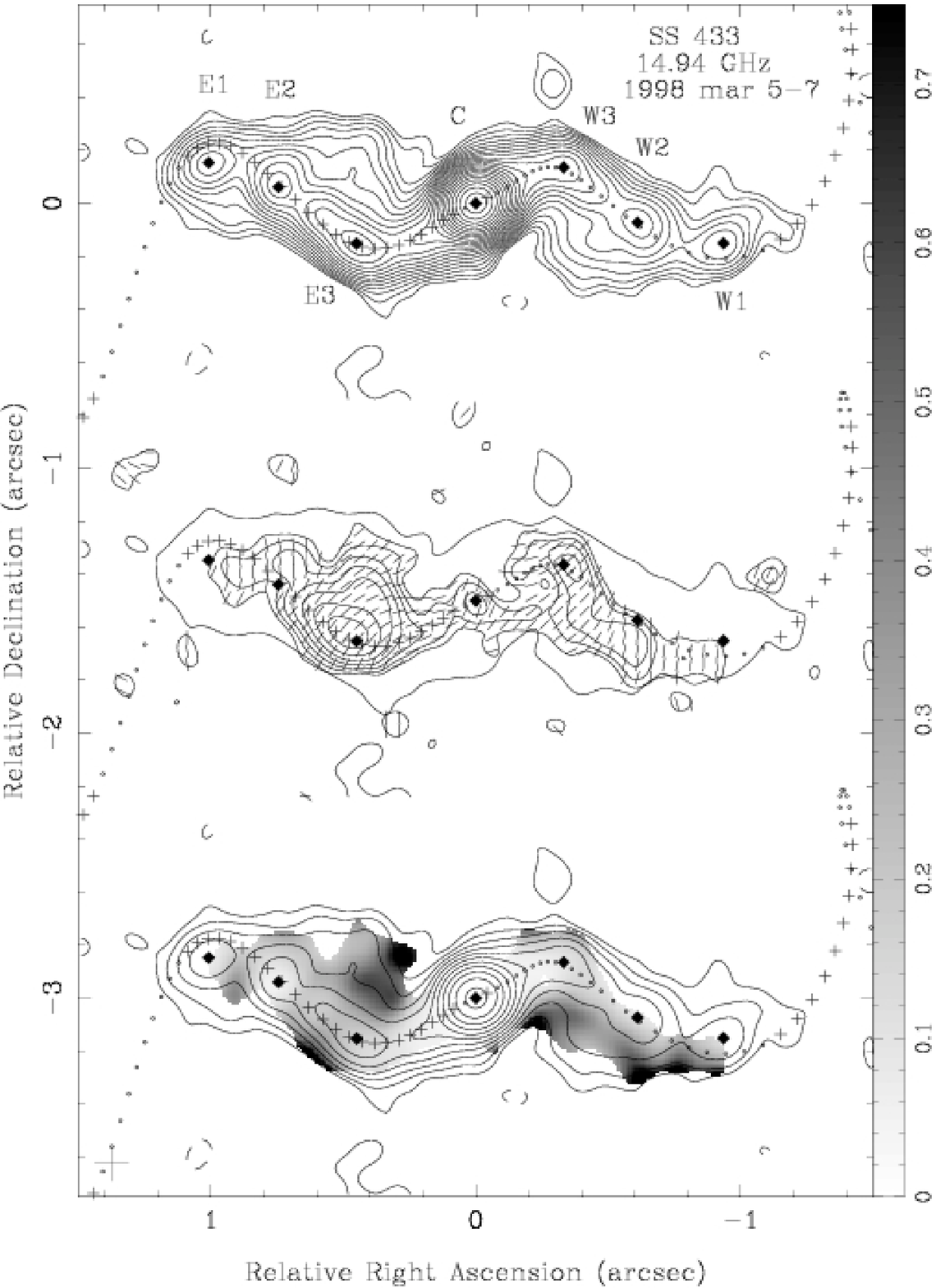}{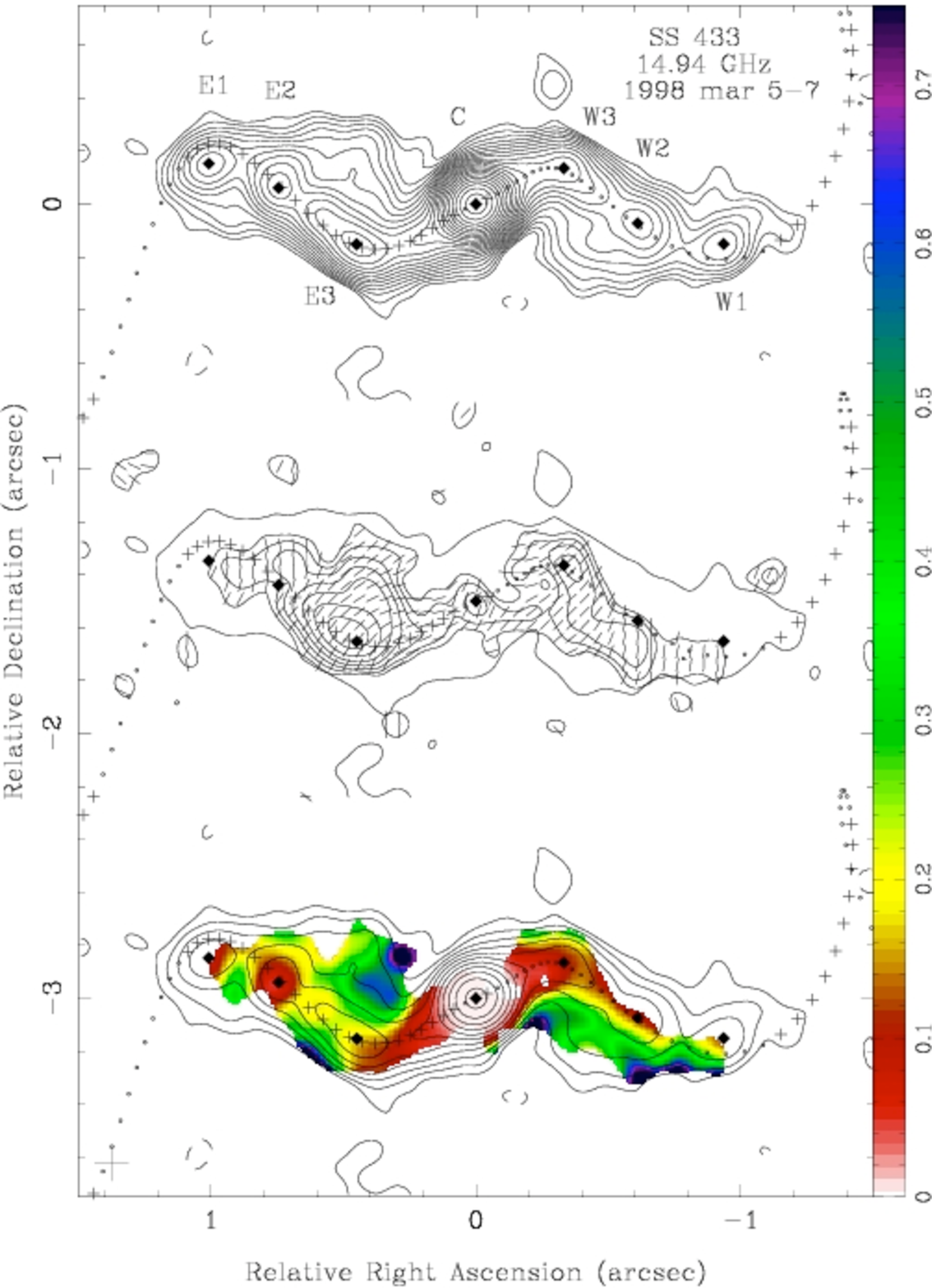}
\caption[]{VLA images of SS\,433 at 14.94~GHz, made from data taken over 1998 March 5-7. Same parts as Figure~\ref{fig:L1}. The contours of total intensity start at 0.15~mJy~beam$^{-1}$, have steps of factors of $\sqrt{2}$, and a peak of 196~mJy~beam$^{-1}$. The  contours of polarized intensity start at 0.15~mJy~beam$^{-1}$, have steps of factors of $\sqrt{2}$, and a peak of 1.5~mJy~beam$^{-1}$. The restoring beam with FWHM of  $0.125\arcsec \times 0.125\arcsec$ is shown as a cross to the lower-left. (Both monochrome and color figures are included for monochrome printers.)
\label{fig:U}}
\end{figure}

\begin{figure}[t]
\plottwo{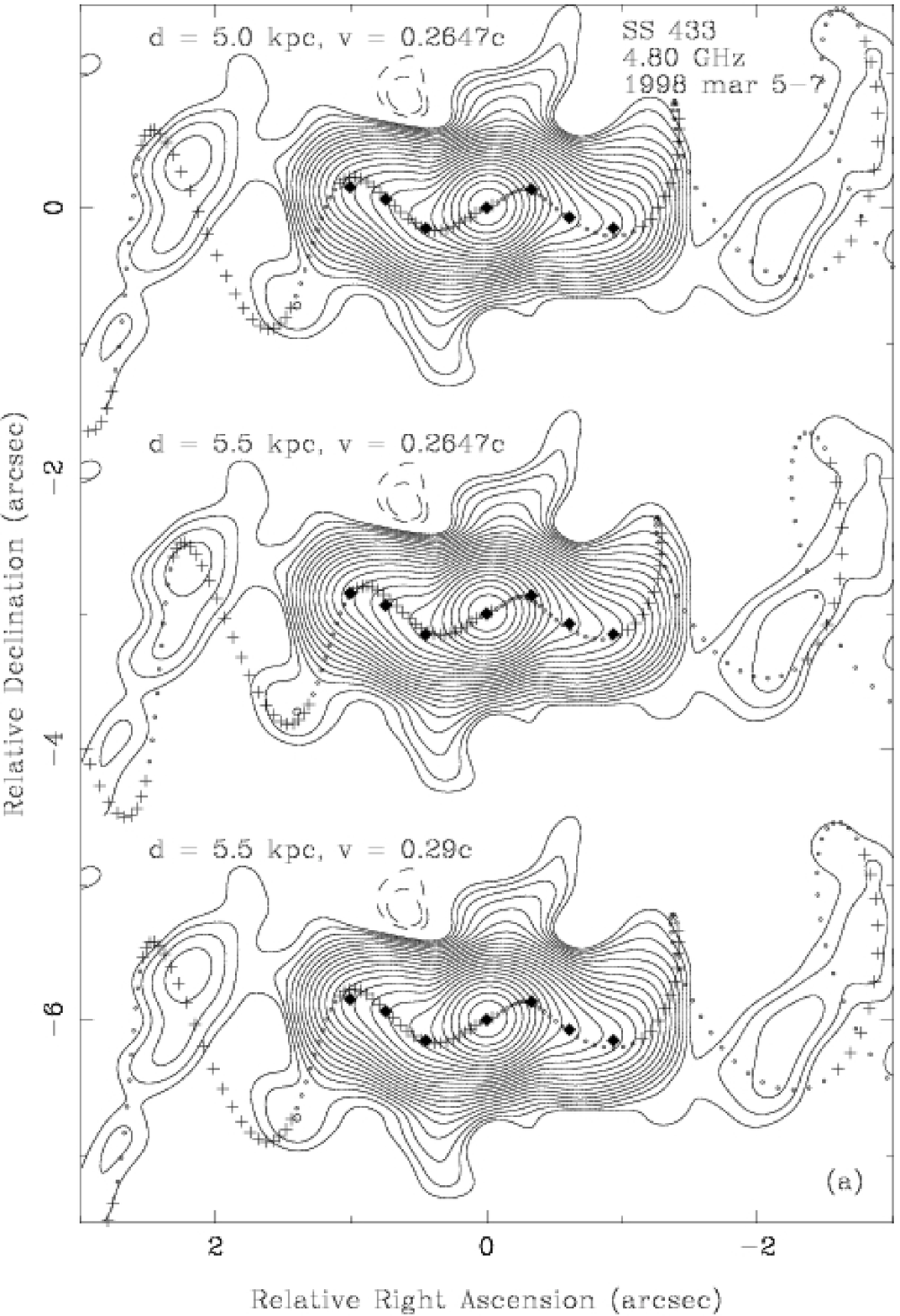}{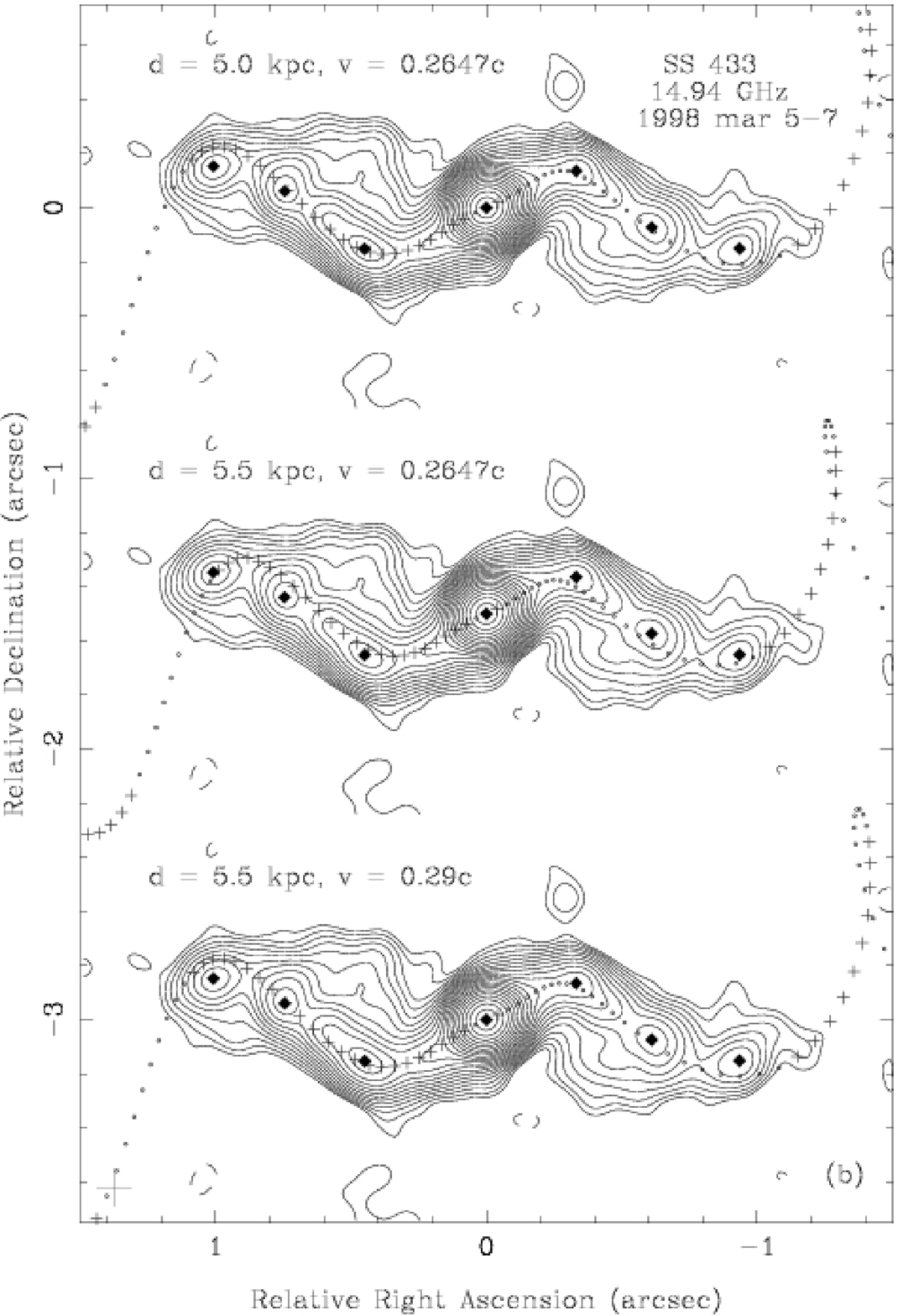}
\caption[]{VLA images of SS\,433 at (a) 4.80~GHz and (b) 14.94~GHz, compared to three kinematic models. In each image we plot models with (top) the optical velocity of $\beta=0.2647$ and distance of $d=5.0$~kpc, (middle) jet with $\beta=0.2647$ and $d=5.5$~kpc, and (bottom) jet with $\beta=0.29$ and $d= 5.5$~kpc.
\label{fig:CU3models}}
\end{figure}

The total intensity structure of SS\,433 at 14.94~GHz (Figure~\ref{fig:U}) shows seven distinct bright spots (the core plus six ``components'') that we denote E1, E2, E3, C, W3, W2, and W1, as labeled on Figure~\ref{fig:U}, and as indicated by dots on the other images. We chose those locations to study the properties of the jets in SS\,433. The 14.94~GHz total intensity image was fit to the components using the AIPS task JMFIT, and the resulting positions used at all frequencies. We measured $I$, $p$, and $\chi$ at each frequency at each component by reading image values at its location. The results for 4.80 and 14.94~GHz are listed in Table~\ref{tab:fluxes}. Since we do not resolve the components cleanly at frequencies other than 14.94~GHz, the corresponding values should be regarded as estimates. The polarization detected at C is sufficiently small that we do not consider it further.

\begin{figure}[t]
\plotone{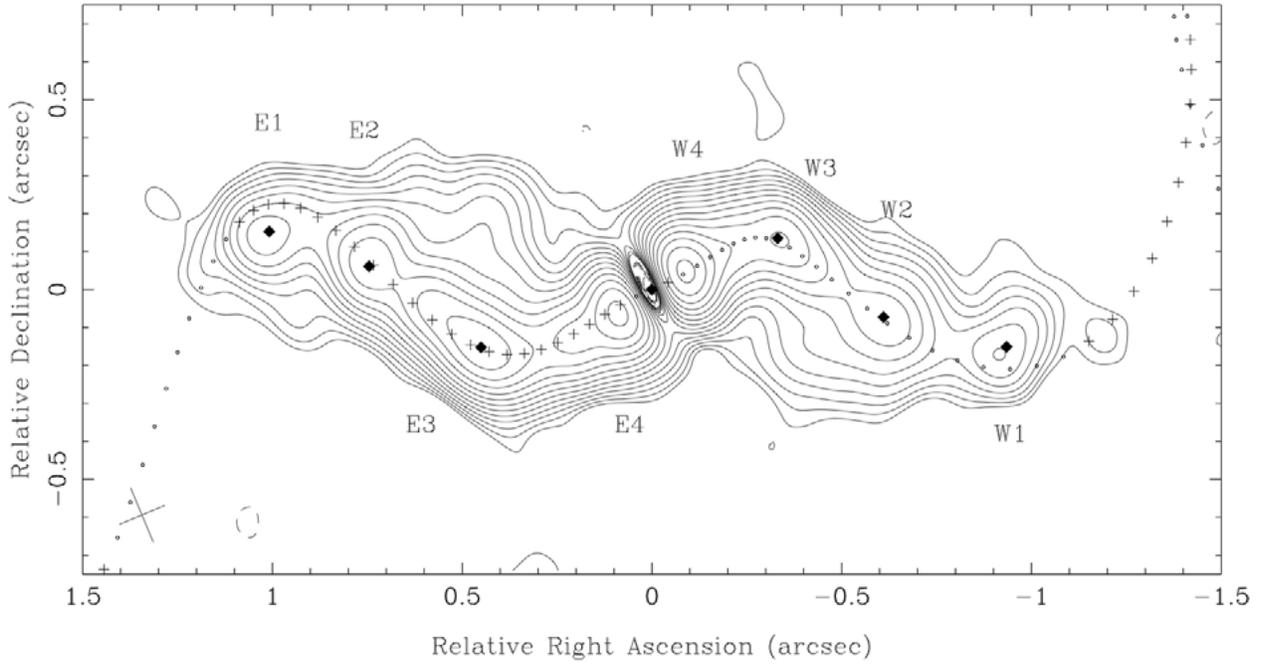}
\caption[]{VLA image of SS\,433 at 14.94~GHz, made from data taken over 1998 March 5-7, with the core removed. The contours of total intensity start at 0.15~mJy~beam$^{-1}$, have steps of factors of $\sqrt{2}$, and a peak of 15~mJy~beam$^{-1}$. The dots indicate the locations of six jet components, as determined from Figure~\ref{fig:U}; the tentative components E4 and W4 found here are also labeled. The kinematic model described in the text is shown as material ejected at five day intervals, with ``+'' and ``o'' indicating oncoming and receding material, respectively. The restoring beam with FWHM of  $0.156\arcsec \times 0.129\arcsec$ is shown as a cross to the lower-left. 
\label{fig:UNC}}
\end{figure}

\begin{figure}[t]
\plotone{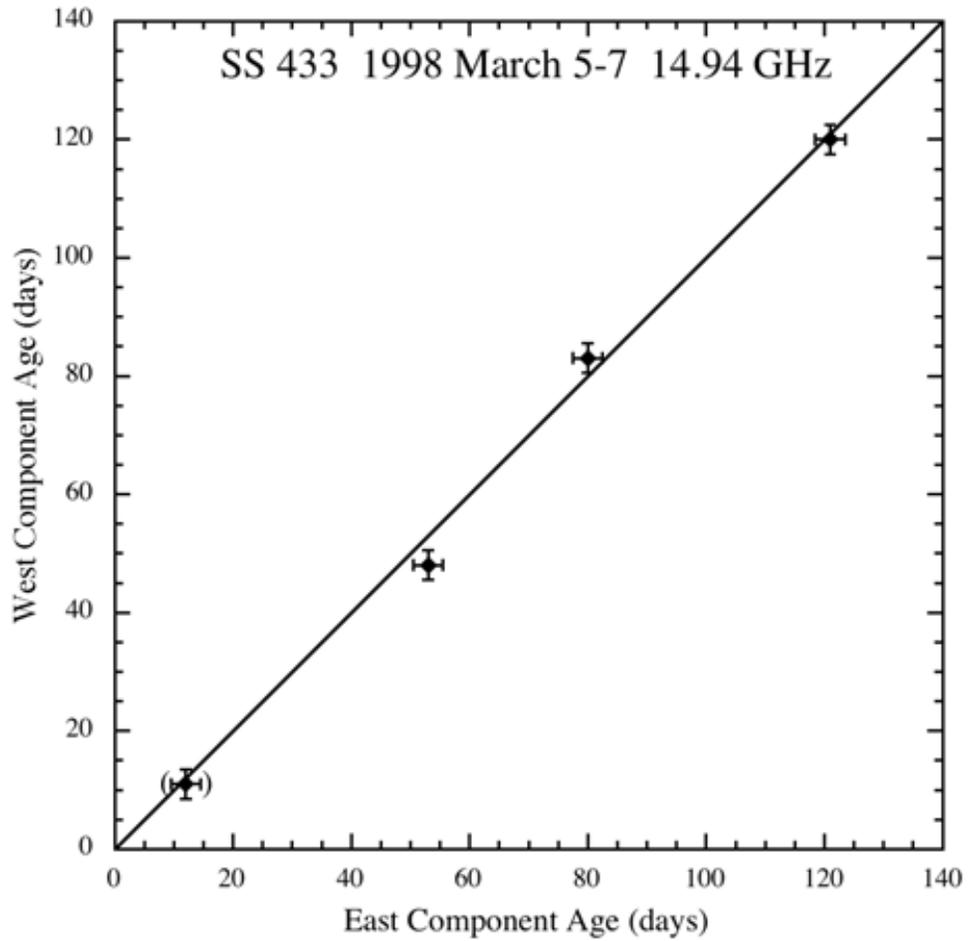}
\caption[]{The model ages of the four pairs of jet components in SS\,433, plotted as age of west component versus age of corresponding east component. The innermost point is not used in the statistical analysis. The sloping line represents the hypothesis that the pairs were emitted simultaneously.
\label{fig:AgePlot}}
\end{figure}

\begin{figure}[t]
\plotone{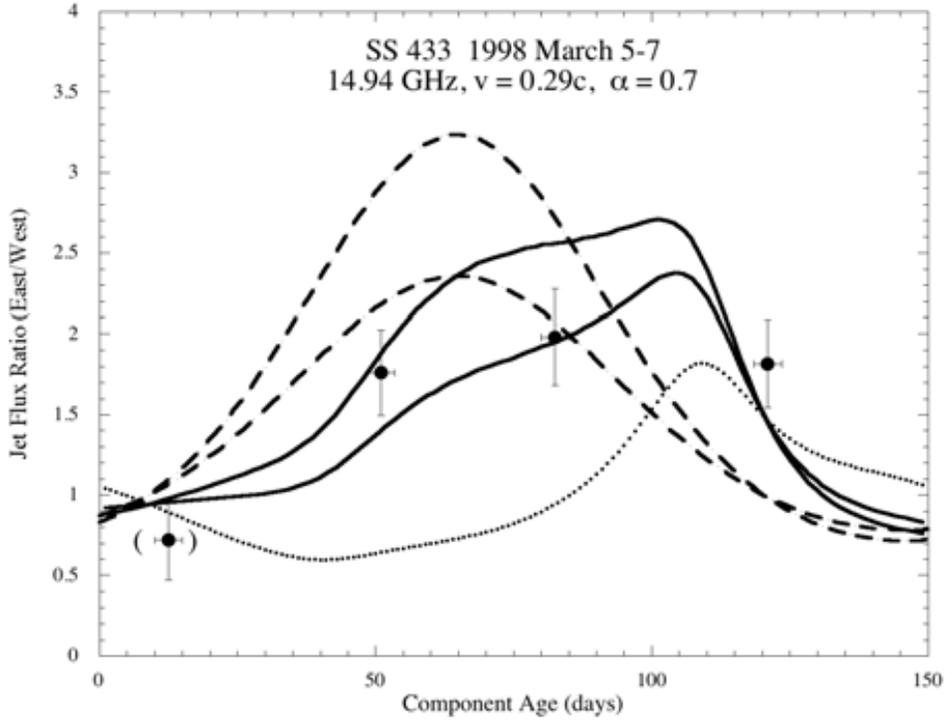}
\caption[]{Predicted versus measured flux ratios in SS\,433 at 14.94~GHz for four pairs of components in the jet of SS\,433, plotted as a function of the average age of each pair of components. Model predictions are based on $d=5.5$~kpc, $v=0.29c$, and $\alpha=0.7$, illustrating (a) projection effects (dotted line), (b) Doppler boosting for a continuous jet (lower dashed line), Doppler boosting for discrete components (upper upper dashed line), and (c) these two effects combined (lower and upper solid lines, respectively).
\label{fig:FluxRatios}}
\end{figure}

\begin{figure}[t]
\plottwo{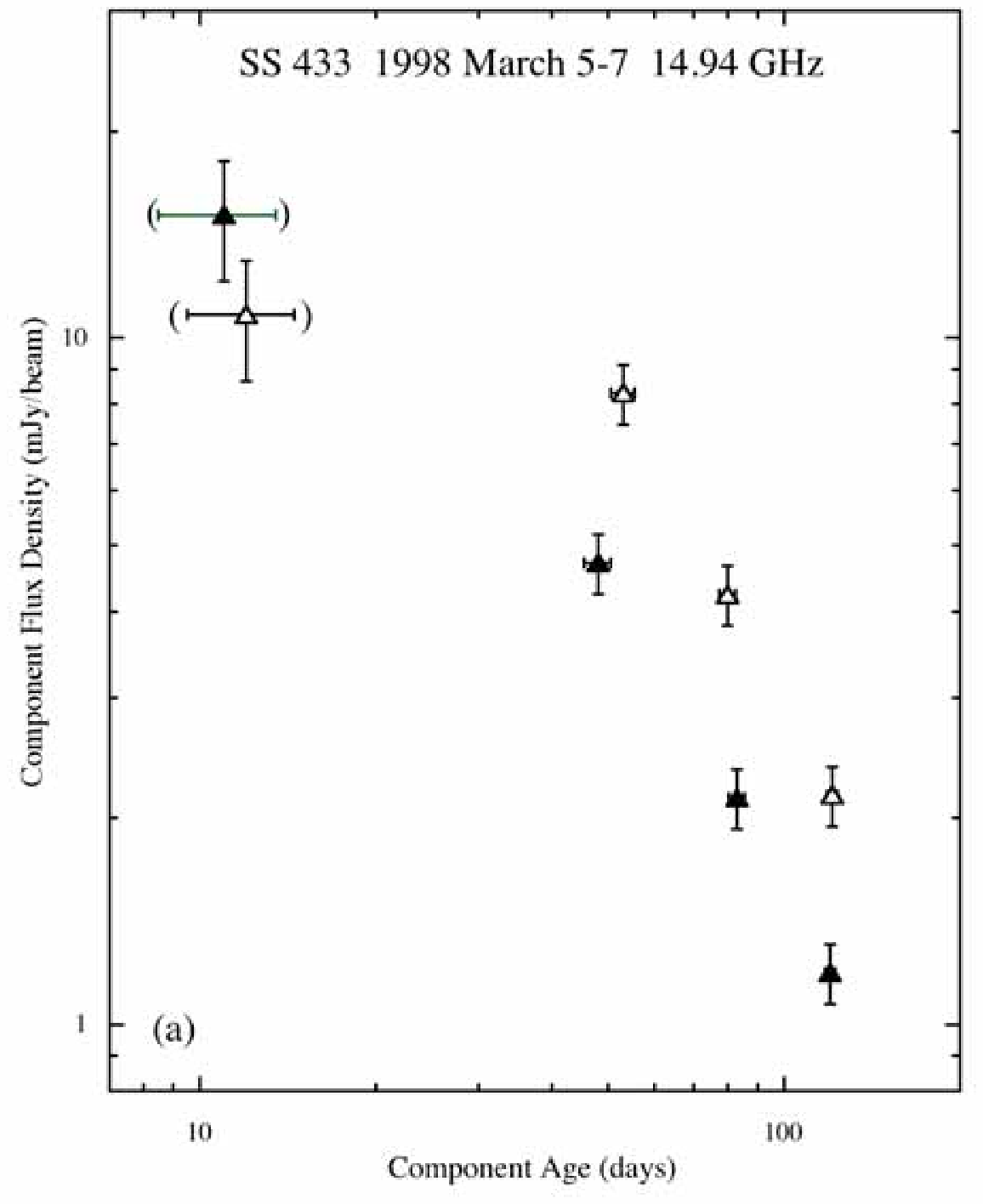}{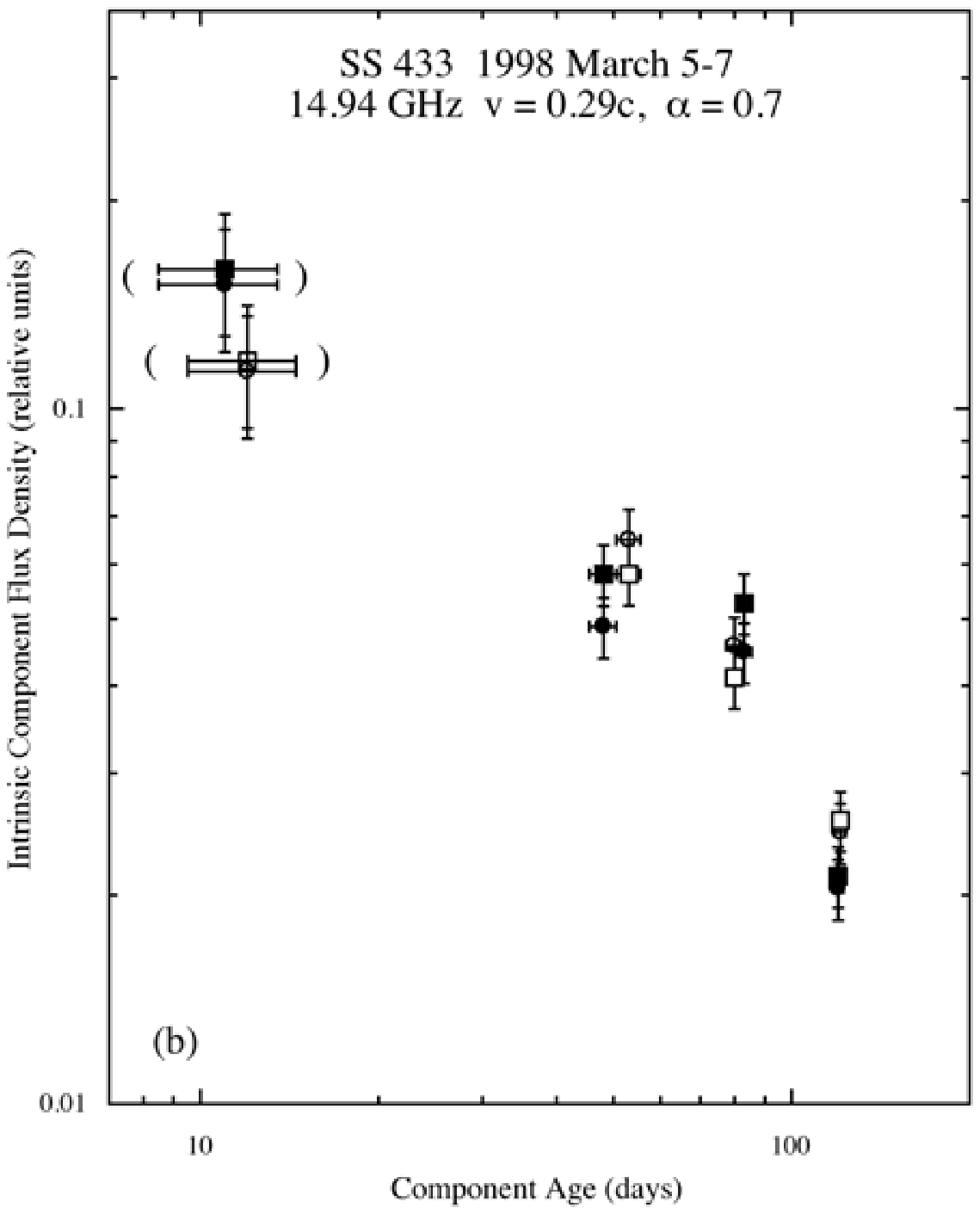}
\caption[]{(a) Observed and (b) intrinsic jet fluxes in SS\,433 at 14.94~GHz for four pairs of components in the jet of SS\,433, plotted as a function of the age of the components. In both parts (a) and (b), components in the east jet are shown as open symbols, those in the west jet as filled symbols. In (b), models for a continuous jet are shown as circles, those for isolated components as squares.
\label{fig:IntrinsicJetFluxes}} 
\end{figure}

We can use the kinematic model to estimate the birth epochs of the six components. Assuming that the precession rate of the jets does not vary while the velocities of ejections may, for components E1--E3 and W1--W3 we used their position angles with respect to the core to find their ages. For E2, E3, W2, \& W3 the model points for $\beta=0.29$ and $d=5.5\mbox{ pc}$ lie on the observed components. For E1 and W1, the model predicts that the components should be further from the core than they are. We attribute this to a smaller ejection speed of $\beta \simeq 0.26$ (essentially the optical value) at the births of these components, which is also the speed needed to fit the larger-scale structure at 4.80~GHz at the locations of the first loops in the jet. The resulting ages are given in Table~\ref{tab:fluxes}. In addition, Figure~\ref{fig:U} contains hints of two components born later than E3 and W3. This is better seen in a 14.94~GHz image from which the core has been removed, shown in Figure~\ref{fig:UNC}. We denote these tentative components E4 and W4. These components are separated from the core by only 0.7--0.9 beamwidths and are much fainter than the core, so we do not include them in the statistical analysis; however, they are included in Figures \ref{fig:AgePlot}--\ref{fig:IntrinsicJetFluxes}. Because their position angles are quite uncertain, the ages of E4 and W4 were estimated using their distances from the core, with the results $\tau\mbox{(E4)} \simeq 12\mbox{ days}$ and  $\tau\mbox{(W4)} \simeq 11\mbox{ days}$. The model ages of the east and west components of each of the four pairs are plotted against each other in Figure~\ref{fig:AgePlot}. It is clear that within the errors, the east and west components of each pair were emitted simultaneously. The intervals between emission of the components on a given side range between 27 and 41 days, with a mean of $35 \pm 3$~days, but the events do not appear to be strictly periodic. Including components E4 and W4 does not alter these conclusions. We note that the 15~GHz VLA image of \cite{Stirl04}, made at precession phase close to ours ($\phi = 0.97$), contains jet peaks with similar intensities and spacings to the ones in Figure~\ref{fig:U} (see their figure 8). It is interesting that \cite{Fiedler87} found a similar characteristic timescale for flaring in SS\,433 of $25 \pm 10\mbox{ days}$.

At 14.94~GHz, we are able to resolve across the jets, and see that the bright ridge-lines are ``leading edges'' of both jets, and are significantly sharper than the trailing edges (Figure~\ref{fig:U}). There is a significant off-jet radiation, most notably north of the jet ridgeline to the east, and south of the ridgeline to the west. Most, but not all, of this emission is within the cone of precession of the jets. This extended plasma may be shed from the jets, or ambient medium in which relativisitic electrons are energized by interaction with the jets.

It is interesting to determine whether or not the east and west jets are intrinsically identical. We can predict the run of brightness down the jets using the kinematic model, including both projection and Doppler effects. In particular, we can calculate the expected ratio of east to west jet brightness, and compare that to the ratios observed in the pairs of peaks E1 \& W1, E2 \& W2, and E3 \& W3. We resolve the components from each other and from the core only at 14.94~GHz, so we consider only that frequency. In all cases we adopt $d=5.5$~kpc and a velocity of $v=0.29c$. First, assuming that it is optically thin, the observed brightness of the jet depends on projection effects, essentially how much of it is included in a resolution element. We predicted the local density of jet material on the sky from the kinematic model by simply counting the number of model points per beam area along the length of the jets. The resulting prediction for the ratio of east jet to west jet brightness (for pieces of the jet of equal ages) as a function of the age of the components is shown in Figure~\ref{fig:FluxRatios}, along with the observed intensity ratios. Second, the model Doppler factors $D$ along the jet were used to predict the flux ratios. For a continuous jet the expected Doppler boost is $D^{2+\alpha}$, while for discrete components it is $D^{3+\alpha}$, where $\alpha \simeq 0.7$ is the jet spectral index (for $F \propto \nu^{-\alpha}$;  \cite{HJ88}). The flux ratios predicted for both a continuous jet and for discrete components are also shown in Figure~\ref{fig:FluxRatios}. We see that both projection effects and Doppler boosting vary slowly down the jet, and that the two are anti-correlelated. Neither projection effects nor Doppler boosting alone can account for the measurements; for example, the former predict that E2 should be fainter than W2 and E3 fainter than W3, while the opposites are true. The combination of projection effects and Doppler boosting produces surprisingly good agreement with the observed ratios, especially for the model of a continuous jet. We regard this as evidence that the two jets are intrinsically very similar, and that both projection effects and Doppler boosts play roles in determining the observed intensity of the jets. More specifically, it confirms that the velocity of the jet material remains relativistic far down the jets from the optically-emitting region, consistent with what is required for the kinematic model to fit the larger-scale radio structure. Since $\beta$ is relatively small in SS\,433, this analysis is not very sensitive to its exact value.

Turning this analysis around, we can investigate the run of intrinsic brightness down the jet by removing both projection effects and Doppler boosting as described above. Observed flux densities and intrinsic brightness (in arbitrary units) are displayed against component ages in Figure~\ref{fig:IntrinsicJetFluxes}. Consistent with the results above, we find that the two jets are intrinsically very similar, and that their brightness falls off with age (or distance from the core), roughly as a power law with $I \propto t^{-1}\propto r^{-1}$, or exponentially with characteristic time scale of about 75~days.

In the conical sheath jet model of \cite{HJ88}, jet intensity falls off with distance down the jet as $I \propto r^{-n}$ with $n = (7p-1)/(6+6\delta)$, where $p$ is the electron energy distribution exponent, and $\delta=0$ and $\delta=1$ represent the freely-expanding  and slowed-expansion cases, respectively. For $\alpha = 0.7$, $p= 2.4$, and our mean index of $n \simeq 1$ corresponds to $\delta \simeq 1.6$, suggesting that the jets are significantly laterally confined as they propagate.

\subsection{Polarization}

The images in Figures~\ref{fig:L1}--\ref{fig:U} show that the jets contain significant polarized fluxes at all frequencies. The fractional polarizations of the six jet components range from a few tenths of a percent at the three lower frequencies, up to $m \simeq 17\%$ for E3 at 14.94~GHz. Along the jet but between the components, the fractional polarization is higher still, reaching $m \simeq \mbox{ 20\%}$ or more (see \S\ref{s:MagneticFields} below).

The off-jet material is highly polarized; at the highest frequency, the fractional polarization is as great as $m \simeq \mbox{ 50 \%}$ (see Figure~\ref{fig:U}). This polarization cannot be due to electron scattering of photons from the core because the observed electric vectors are not circumferential. We are left to interpret the large $m$ as due to synchrotron radiation where there is significant ordering of the magnetic field. This strong polarized signal may confuse the determination of magnetic field orientations in the jet itself (see \S\ref{s:MagneticFields} below).

The small fractional polarization seen in the jets at the low frequencies is probably due either to side-side (beam) depolarization, or front-back depolarization consistent with large internal Faraday rotation. The east jet is significantly more highly polarized than the west jet at the low frequencies; this may be because the east jet is usually closer to us than is the west jet, and therefore suffers less depolarization near the source.

\subsection{Faraday Rotation}

Because SS\,433 has baryonic jets and is immersed in a stellar wind, we expect that Faraday rotation could be important. We examined this first by plotting the electric vector position angles $\chi$ as a function of wavelength for each component at the three lower frequencies. Comparison of the middle panels of Figures~\ref{fig:L1}--\ref{fig:L3} shows a clear electric vector position angle rotation with frequency. In the case of a foreground Faraday-rotating screen, the observed electric vector position angles, the wavelengths, and the rotation measure RM are related by $ \chi = \chi_0 + \mbox{RM}~\lambda^2,$ where $\chi_0$ is the underlying (unrotated) electric vector orientation. The results are consistent with an average foreground rotation measure of $\sim$110$\mbox{ rad m}^{-2}$, in agreement with the results of \cite{Stirl04}.

To examine the Faraday rotation distribution on smaller scales we derived rotation measures at each component using only  the 4.80~GHz image and the convolved 14.94~GHz image. At frequencies of 4.80 and 14.94~GHz, a rotation measure of $\mbox{RM} = 100$~rad~m$^{-2}$ corresponds to rotations $\Delta\chi$ of 22$^{\circ}$ and 2.3$^{\circ}$, respectively, and the $180^{\circ}$ ambiguity in $\chi$ corresponds to $\Delta$RM = $898~\mbox{rad m}^{-2}$. With only two frequencies, were are unable to determine unambiguous rotation measures; nonetheless, we can estimate the size of the Faraday effect at each component by fitting the two frequencies to a $\lambda^2$-law, and choosing the position angle ambiguity that leads to the smallest RM. It is clear that the magnitude of the rotation measure at W3 is much larger than at the other components, and for it we determined the two smallest possible rotation measures magnitudes. This yields the rotation measures and intrinsic electric vector position angles in Table~\ref{tab:RM}. 

\begin{figure}[t]
\plotone{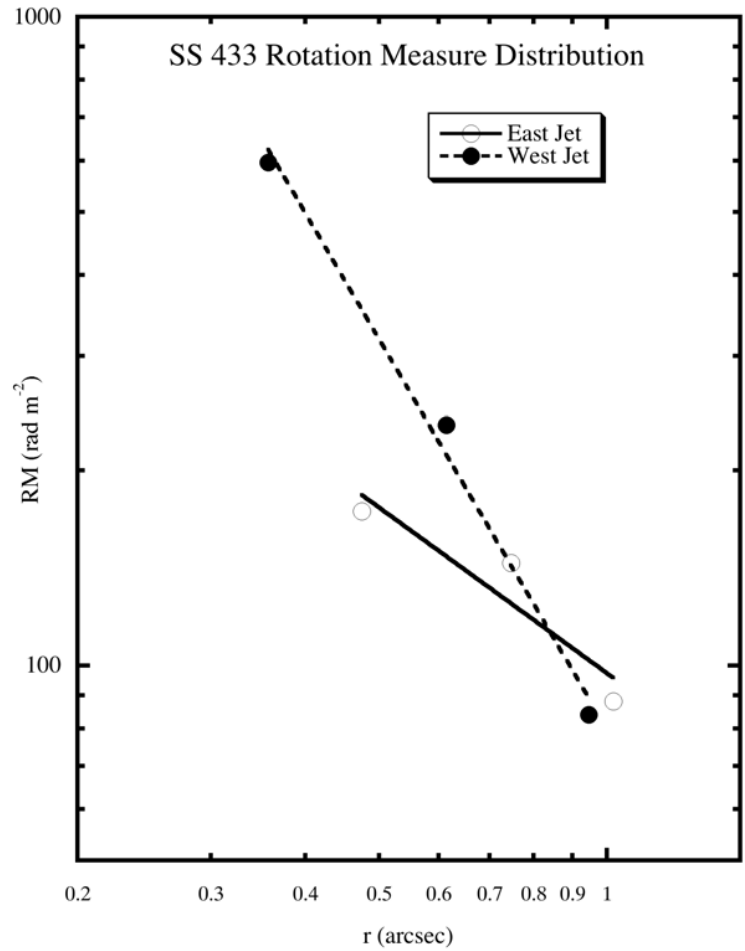}
\caption[]{Rotation measures in SS\,433 as a function of radial distance from the core, for the six bright jet components described in the text. Shown is the rotation measure determined from the 4.80 and convolved 14.94~GHz images. The power law fits are described in the text.\label{fig:RM-PLOT}}
\end{figure}

A plot of Faraday rotation versus radial distance is shown in Figure~\ref{fig:RM-PLOT}. It is consistent with power law fall-off $\mbox{RM} \propto r^{-n}$, with indices of $n \simeq 0.8$ for the east jet and $n \simeq 2$ for the west. Multi-frequency multi-epoch VLA observations currently being reduced should show if and how this changes as the jets precess. We note that \cite{Stirl04} also detected rotation measure gradients across SS\,433, and that their values are somewhat different than ours. The increase in rotation measure towards the core probably accounts for its negligible polarization through Faraday depolarization.

\subsection{Magnetic Fields}
\label{s:MagneticFields}

\begin{figure}
\plotone{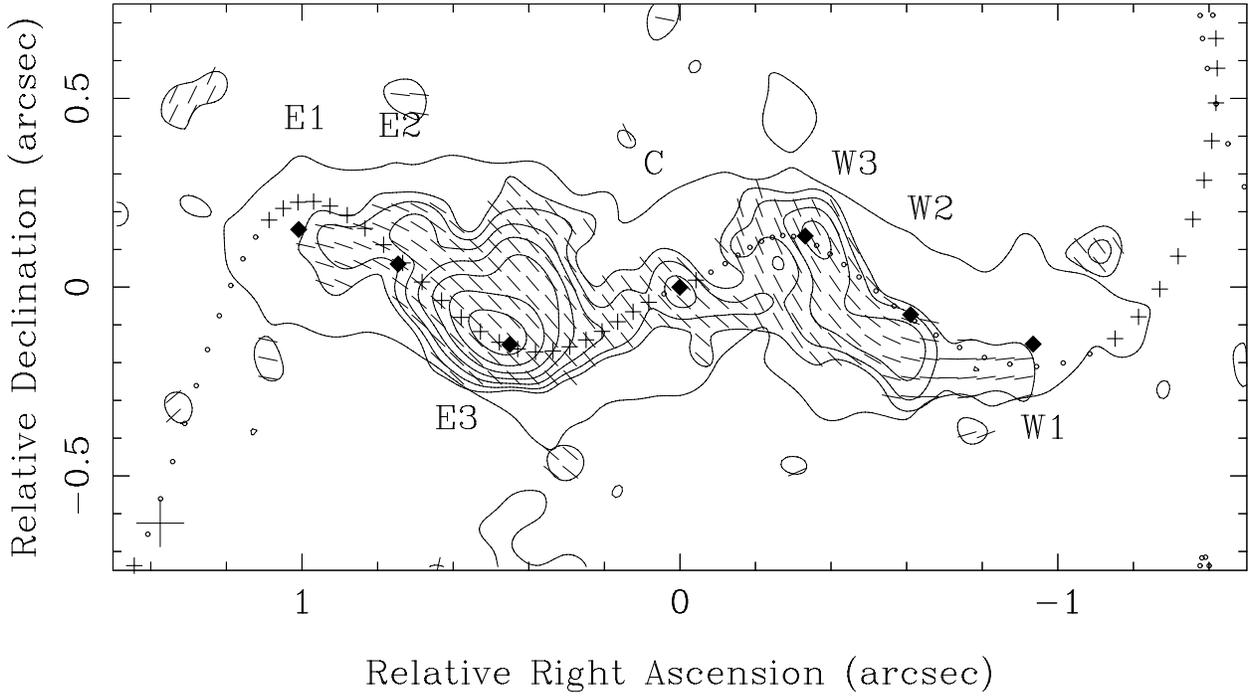}
\caption[]{Magnetic field orientations in SS\,433. The contours are polarization intensity, with tick marks of Faraday-rotation-corrected magnetic field vectors, and a single total intensity contour at 0.15~mJy~beam$^{-1}$. The polarization contours start at 0.15~mJy~beam$^{-1}$, have steps of factors of $\sqrt{2}$, and a peak of 1.5~mJy~beam$^{-1}$.  The kinematic model described in the text is shown as material ejected at five day intervals, with ``+''  and ``o'' indicating oncoming  and receding parts, respectively. The restoring beam with FWHM of  $0.125\arcsec \times 0.125\arcsec$ is shown as a cross to the lower-left. \label{fig:Bplot}}
\end{figure}

Interpreting the radio emission as synchrotron radiation, the magnetic field vectors projected on the sky are perpendicular to the Faraday-rotation-corrected electric vectors $\chi_0$. The distribution of inferred magnetic field vectors derived in this way is shown in Figure~\ref{fig:Bplot}. Along the jets from E3 to E1 and from W3 to W1, the magnetic field directions curve gently with the jets, and lie roughly tangent to the ridge-lines. They are not aligned with the direction of ballistic motion of the jet material. This is consistent with the results of \cite{Stirl04}.  The regions near E1 and W1 are harder to interepret because the jet trajectory is strongly curved and the polarized signal-to-noise ratio is rapidly decreasing. Note also that in the region E3--C--W3 the inferred magnetic field is roughly orthogonal to the jet locus. However, the fractional polarization there is small, and we do not regard this result as secure.

The presence of highly-polarized off-jet emission is a surprise (high polarization of off-jet material is hinted at in figure~4 of \cite{Stirl04}). In the regions north of E3 and south of W3 the magnetic field is remarkably unidirectional, lying at a position angle $\chi \simeq 45^{\circ}$. These high fractional polarizations show that the magnetic fields in the ambient medium are ordered on scales of thousands of AU. One possibility is that interactions between the jets and the ambient medium result in large scale motions of the surrounding gas that order an initially tangled field by shearing it. We do not know the three-dimensional distribution of this material, but it does appear to be trailing the jets, and thus probably lies inside the precession cone or on its boundary.

The strongly polarized off-jet emission confuses the determination of the magnetic field orientations in the jet helices themselves. The jet ridge-lines are loci of minimum fractional polarization, with $m$ higher in the adjacent off-jet regions. This suggests that the jet polarization is contaminated by radiation from foreground and/or background ambient medium. We suggest that  the jet magnetic fields are everywhere parallel to the observed helix. Between E3 and E2 and between W3 and W2 the jet emission and the off-jet emission have the same magnetic field orientation. If the fractional polarization of the jet itself is smaller than that of the surrounding medium, a local minimum of $m$ will be created along these parts of the jets. Between C and E3 and between C and W3, the jet and off-jet polarizations are orthogonal, with the off-jet radiation winning out, and the apparent polarization of the jet is lower than that of the adjacent emission. The polarized signals in the regions between E2 and E1 and  between W2 and W1 are weaker, but they appear be consistent with magnetic fields aligned with the helix.

At 14.94~GHz we can reliably determine the fractional polarization of the inter-component regions in the east jet, between E1 and E2 and between E2 and E3. From Figure~\ref{fig:U} it is clear that $m$ is larger in the these regions than in the adjacent components. In the west jet the same pattern occurs at lower signal-to-noise ratio. This suggests that the jets contain some tangled field that is partially ordered into a transverse orientation by compressions that create the components. This process is apparently present in knots in parsec-scale jets of some AGN \citep{HAA}. One extragalactic jet in which the order in a predominantly longitudinal field is apparently modified in this manner is 3C\,345 \citep{BRW94,WCRB94}. If the components in SS\,433 are the loci of moving shocks, then some of the analysis in \S\ref{s:structure} might have to be modified; however, we know that the bulk velocity in the jets cannot vary significantly as they flow outward or else the kinematic model would fail to fit the structure beyond about $1^{\prime\prime}$. Only measurement of the proper motions of jet components can determine if they are moving at the jet fluid speed.\footnote{Unfortunately, the ongoing upgrade of the VLA is resulting in EVLA antennas without 15~GHz receiver systems, a condition that will be remedied eventually.}

Magnetic field lines that are tangent to the jet ridgeline and the strong leading-edge polarization suggest that the field orientations are determined at least in part by the interaction of the jet with the surrounding medium. For example, this is consistent with a model in which jet material plows into an ambient medium containing tangled field, and the resulting compression orders the field in two dimensions \citep{Laing}. Because of the precession of the jets, this process would occur predominantly on the ``outsides'' of the helices, and produce the sharp, highly-polarized leading edges that we see. Alternatively, shear created at jet-ambient medium interface could stretch a tangled field into a longitudinal one.

\section{CONCLUSIONS}
\label{s:conclude}

We have made polarization-sensitive VLA A-array images of SS\,4333 at five frequencies from 1.4 to 15~GHz, using archival data over three days from March 1998. The principle results in this paper are:

\begin{enumerate}

\item The structure of the SS\,433 jets at 4.80 and 14.94~GHz is consistent with the kinematic model and with a distance of 5.5~kpc if the velocities of ejection vary by about $\pm$10\% around the optically determined value.

\item The 14.94~GHz image of the jets reveals a continuous jet, on top of which there are four
pairs of brighter components. Comparison with the kinematic model shows that the two members of each pair were emitted simultaneously and with equal velocities. Pairs emerge at intervals of about $35\mbox{ days}$, but the process is not strictly periodic.

\item When the observed brightness ratios of the pairs of components are corrected for projection effects and Doppler boosting, we find that the two jets are intrinsically very similar.

\item The intrinsic intensity of the jet components falls of with their age (or distance from the core) roughly as a power law with index $-1$, or exponentially with a time scale of about 75~days.

\item The jets are linearly polarized with fractional polarization reaching $m \simeq 20\%$, consistent with synchrotron radiation from relativistic electrons. The core is essentially unpolarized, probably due to Faraday depolarization.

\item The Faraday rotation is consistent with a foreground screen with RM~$\simeq110 ~ \mbox{rad m}^{-2}$, plus a larger contribution of several hundred rad~m$^{-2}$ close to the core that is asymmetrical across the source.

\item The jets are surrounded by significant off-jet emission that is primarily trailing the jet ridgeline.
This emission is highly linearly polarized, ranging as high as $m \simeq 50\%$ or more, indicating the presence of large-scale magnetic fields and relativistic electrons in the ambient medium. The jet ridgeline is a locus of minimum fractional polarization, suggesting contamination by foreground and/or background off-jet emission, and confusing the determination of the field orientation in the helices themselves. Taking this into account, the polarization images are consistent with a magnetic field that is everywhere parallel to the jet locus.

\item The fractional polarization of the jets is greater between the components than in the components, suggesting that the jet contains some tangled magnetic field that is partially ordered by compressions at the components, as seen in extragalactic jets.

\item The fact that the leading edges of the jets are sharper in profile than the trailing edges and the longitudinal orientation of the magnetic field show that significant interactions between the jets and the surrounding medium play an essential role in determining the morphology and magnetic field of SS\,433 jets on the scales seen in VLA images.

\end{enumerate}
Further polarization-sensitive observations covering a range of precession phases and at multiple frequencies that are currently being analyzed should shed more light on the role of jet-ambient medium interactions in SS\,433. This may lead to new perspectives on similar phenomena taking place in the parsec-scale jets of  AGN.

\acknowledgments

This material is based upon work supported by the National Science Foundation 
under Grant No.~0307531, and prior grants. Any opinions, findings, and conclusions 
or recommendations expressed in this material are those of the authors and do not necessarily reflect  the views of the National Science Foundation. The National Radio Astronomy Observatory is a facility of the National Science Foundation, operated under cooperative agreement by Associated Universities, Inc. We thank Tingdong Chen and Teddy Cheung for their assistance with AIPS and DIFMAP, Dan Homan for his plotting scripts, and Vivek Dhawan, Amy Mioduszewski, and Michael Rupen for stimulating our interest in this subject. The anonymous referee provided helpful comments that led to clarification of the manuscript.

Facilities: \facility{VLA (A array)}

\clearpage


\begin{deluxetable}{lllc}

\tablecaption{Ephemeris for SS\,433.\label{tab:ephem}}
\tablewidth{0pt}
\tablecolumns{4}
\tablehead{\colhead{Property} & \colhead{Symbol\tablenotemark{a}} & \colhead{Value} & \colhead{Reference\tablenotemark{b}} }

\startdata

Jet Speed & $\beta = v/c$ & $0.26\pm10\%$ & 1 \\
Precession Period & $P$ & 162.375~d  & 2 \\
Reference Epoch (TJD) & $t_{ref}$ & 3563.23 & 2 \\
Precession Cone Opening Angle & $\psi$ & $20.92^{\circ}$ & 2 \\
Precession Axis Inclination & $i$ & $78.05^{\circ}$ & 2 \\
Precession Axis Position Angle & $\chi + \pi/2$ & $98.2^{\circ}$ & 3 \\
Sense of Precession & $s_{rot}$ & $-1$ & 4 \\
Distance & $d$ & $5.5~\mbox{kpc}$ & 5  \\ 

\enddata

\tablenotetext{a}{Following \cite{HJ81b}.}
\tablenotetext{b}{References--(1) See text,~(2) \cite{Eik},~(3) \cite{Modu},~(4) \cite{HJ81b}, ~(5) \cite{BB04,LBG07}.}

\end{deluxetable}

\clearpage

\begin{deluxetable}{rrrrr}

\tablewidth{0pt}
\tablecolumns{5}
\tablecaption{Components in the Jets of SS\,433, Epoch 1998 March 5-7.\label{tab:fluxes}}
\tablehead{\colhead{$\nu$(GHz)} & \colhead{$I$(mJy)} & \colhead{$p$(mJy)} & \colhead{$m$(\%)} & \colhead{$\chi$} }

\startdata


\cutinhead{E1: $(+1.008,+0.153)^a$, $(1.020, 81^{\circ})^b$, $\tau = 121^c$}
4.800 & \phn16.5\phn  & 1.83\phn   & 11.1 & \phn$+9^{\circ}$ \\
14.94\phn & \phn\phn2.16   & 0.140   & \phn6.5 & $-22^{\circ}$ \\ 
14.94$^d$    & \phn\phn4.68    & 0.225    & \phn4.8  & \phn{$-8^{\circ}$}  \\ 
\cutinhead{E2: $(+0.745,+0.061)^a$, $(0.747, 85^{\circ})^b$, $\tau = 80^c$}
4.800         & \phn34.1\phn & 4.93\phn & 14.8 & \phn$-1^{\circ}$ \\
14.94\phn & \phn\phn4.24 & 0.245 & \phn5.8  & \phn$-7^{\circ}$ \\ 
14.94$^d$    & \phn\phn10.4\phn & 1.08\phn & 10.4  & $-30^{\circ}$ \\ 
\cutinhead{E3: $(+0.450,-0.152)^a$, $(0.475, 109^{\circ})^b$, $\tau = 53^c$}
4.800 & \phn59.9\phn & 6.39\phn & 10.7  & $-7^{\circ}$ \\
14.94\phn & \phn8.31 & 1.43\phn & 17.2 & $-37^{\circ}$ \\ 
14.94$^d$    & \phn20.1\phn & 2.55\phn & 12.7 & $-42^{\circ}$ \\ 
\cutinhead{C: $(0,0)^a$, $(0,0)^b$}
4.800         & 398.\phn\phn   & 0.840       & \phn0.2     & $-55^{\circ}$ \\
14.94\phn & 196.\phn\phn    & 0.349       & \phn0.2     & $-15^{\circ}$ \\ 
14.94$^d$       & 214.\phn\phn    & 0.518   & \phn0.2     & $-47^{\circ}$ \\ 
\cutinhead{W3: $(-0.331,+0.135)^a$, $(0.357, -68^{\circ})^b$, $\tau =48^c$}
4.800         & \phn78.0\phn   & 1.31\phn    & \phn1.7 & $+70^{\circ}$ \\
14.94\phn & \phn\phn4.72    & 0.431         & \phn9.1 & $-46^{\circ}$ \\ 
14.94$^d$ & \phn26.2\phn    & 0.745         & \phn2.8 & $-49^{\circ}$ \\ 
\cutinhead{W2: $(-0.610,-0.073)^a$, $(0.614, -97^{\circ})^b$, $\tau = 83^c$}
4.800             & \phn21.6\phn    & 0.294          & \phn1.4  & $+31^{\circ}$ \\
14.94\phn     & \phn\phn2.14    & 0.217          & 10.1   & \phn$-4^{\circ}$ \\ 
14.94$^d$     & \phn\phn5.88    & 0.390          & \phn6.6   & $-16^{\circ}$ \\ 
\cutinhead{W1: $(-0.934,-0.151)^a$, $(0.946, -99^{\circ})^b$, $\tau = 120^c$}
4.800 & \phn10.8\phn   & 1.08\phn             & 10.0     & $+20^{\circ}$ \\
14.94\phn & \phn\phn1.19   & 0.148      & 12.4     & \phn$-2^{\circ}$ \\ 
14.94$^d$ & \phn\phn2.41   & 0.128      & \phn5.3     & $\phn+3^{\circ}$ \\ 

\enddata

\tablenotetext{a}{Position relative to the core, as $(\alpha,\delta)$ in arcseconds.} 

\tablenotetext{b}{Distance from the core in arcseconds and structural position angle.}

\tablenotetext{c}{Model component age in days.} 

\tablenotetext{d}{Values measured at the resolution of the 4.800~GHz images.} 

\end{deluxetable}

\clearpage


\begin{deluxetable}{cccc}

\tablecaption{Faraday Rotation in SS\,433, Epoch 1998 March 5-7.\label{tab:RM}}
\tablewidth{0pt}
\tablecolumns{4}
\tablehead{ \colhead{Component} & \colhead{RM (rad m$^{-2}$)} & \colhead{$\chi_0$}  
& \colhead{ $\theta_B$\tablenotemark{a} } }

\startdata

E1\phn	&	\phn$+88$	&	$-11^{\circ}$		&	$+79^{\circ}$	\\
E2\phn	&	$+144$	         &	$-33^{\circ}$		&	$+57^{\circ}$	\\
E3\phn	&	$+173$		&	$-46^{\circ}$		&	$+44^{\circ}$	\\
W3$^b$	&	$-302$		&	$-42^{\circ}$		&	$+48^{\circ}$	\\
W3$^b$   	&	$+596$		&	$-63^{\circ}$		&	$+27^{\circ}$	\\
W2\phn	&	$+235$	         &	$-22^{\circ}$	          &	$+68^{\circ}$	\\
W1\phn	&	\phn$+84$&	\phn$+1^{\circ}$	&	$-89^{\circ}$	\\

\enddata

\tablenotetext{a}{Position angle of the underlying magnetic field.}

\tablenotetext{b}{Two possible rotation measures.}

\end{deluxetable}

\clearpage

\clearpage

\clearpage

\clearpage

\clearpage

\clearpage

\clearpage

\epsscale{1.00}

\clearpage

\epsscale{1.00}

\clearpage

\clearpage

\epsscale{1.0}

\clearpage

\epsscale{1.2}
\epsscale{1.0}

\clearpage

\epsscale{0.80}
\epsscale{1.00}

\clearpage

\epsscale{1.0}

\end{document}